\documentclass[%
 reprint,
nofootinbib,
nobibnotes,
 amsmath,amssymb,aps,
 prd,
]{revtex4-2}

\usepackage{graphicx}
\usepackage{dcolumn}
\usepackage{bm}
\usepackage{todonotes}
\usepackage{hyperref}

\makeatletter
\let\Hy@backout\@gobble
\makeatother

\begin{document}


\title{Directed search for continuous gravitational-wave signals from the Galactic Center in the Advanced LIGO second observing run}%
\author{Ornella J. Piccinni$^{1,2}$}
\email{ornella.juliana.piccinni@roma1.infn.it}
\author{P. Astone$^1$}
\author{S. D'Antonio$^3$}
\author{S. Frasca$^{1,2}$}
\author{G. Intini$^{1,2}$}
\author{I. La Rosa$^4$}
\author{P. Leaci$^{1,2}$}
\author{S. Mastrogiovanni$^5$}
\author{A. Miller$^{1,2,6}$}
\author{C. Palomba$^1$}

\affiliation{$^1$INFN, Sezione di Roma, I-00185 Roma, Italy\\ $^2$University of Rome Sapienza, I-00185 Roma, Italy \\ $^3$INFN, Sezione di Roma Tor Vergata, I-00133 Roma, Italy \\ $^4$Laboratoire d'Annecy-le-Vieux de Physique des Particules (LAPP), Universite Savoie Mont Blanc, CNRS/IN2P3, F-74941 Annecy, France\\$^5$PC, AstroParticule et Cosmologie, Universite Paris Diderot, CNRS/IN2P3, CEA/Irfu, Observatoire de Paris, Sorbonne Paris Cite, F-75205 Paris Cedex 13, France\\ $^6$University of Florida, Gainseville, Florida 32611, USA}

\date{\today}

\begin{abstract}
In this work we present the results of a search for continuous gravitational waves from the Galactic Center using LIGO O2 data. The search uses the Band-Sampled-Data directed search pipeline, which performs a semi-coherent wide-parameter-space search, exploiting the robustness of the FrequencyHough transform algorithm. The search targets signals emitted by isolated asymmetric spinning neutron stars, located within 25-150 parsecs from the Galactic Center. The frequencies covered in this search range between 10  and 710 Hz with a spin-down range from $-1.8\times10^{-9}$  to $3.7\times10^{-11}$ Hz/s. No continuous wave signal has been detected and upper limits on the gravitational wave amplitude are presented. The most stringent upper limit at $95\%$ confidence level, for the Livingston detector, is  $\sim 1.4 \times 10^{-25}$ at frequencies near 160 Hz. To date, this is the most sensitive directed search for continuous gravitational-wave signals from the Galactic Center and the first search of this kind using the LIGO second observing run.  
\end{abstract}

\maketitle

\section{\label{sec:level1}Introduction}
Gravitational wave (GW) signals are produced whenever there is a mass quadrupole variation, given for example by fast moving compact objects. All gravitational wave signals detected so far by the LIGO \cite{adligo}  and Virgo \cite{advirgo} interferometers, during the first two observational runs, have a short time duration and have been produced by the coalescence of a pair of Black Holes (BHs) or Neutron Stars (NSs) \cite{O1O2events}.

Beside transient GW signals, also long-lived coherent signals are expected to be found in LIGO-Virgo data. This type of GW is called Continuous gravitational Wave (CW). Astrophysical systems that can emit CWs are, for example, fast spinning galactic NSs, asymmetric with respect to their rotation axis, isolated or in binary systems. Another more exotic source of CWs is ultralight bosons clouds around BHs \cite{bosonsOtto,bosonsArvanitaki,bosonsBrito2017}. A comprehensive review of potential CW sources can be found in \cite{riles,prix2009}. 

Several different mechanisms have been proposed to explain the existence of the star asymmetry which triggers the GW emission  \cite{lasky,glampe}. It can be caused by the presence of elastic stresses, strong internal magnetic fields not aligned to the star rotation axis, free precession with respect to the star rotation axis, excitation of long-lasting r-mode oscillations and the accretion of matter from a companion star, e.g. in Low-Mass X-ray Binaries (LMXB). The degree of asymmetry, usually referred to as \emph{ellipticity}, is strictly connected with the strain the star can sustain, hence to the property of matter inside the star and its equation of state \cite{mcdanielowen,owen2005,Woan2019}.

CW signals are nearly monochromatic with a frequency $f_{GW}$ proportional to the star spin frequency and a duration longer than the observational time (of the order of months or years). The signal arriving at the detector is indeed not monochromatic, since some modulations occur, mainly caused by the source intrinsic spin-down and by the Doppler effect.   

For the prototypical case of an isolated spinning NS, non-axisymmetric with respect to the rotational axis, and located at a distance $d$ from the detector,  the GW-strain amplitude $h_0$ is given by
\begin{equation}
\label{eqn:strain}
 h_0=\frac{4\pi^2G}{c^4} \frac{I_{zz} f_{GW}^2}{d}\epsilon,
\end{equation}
where $I_{zz}$ is the star moment of inertia around the rotation axis ($z$-axis) while $\epsilon=\frac{I_{xx}-I_{yy}}{I_{zz}}$ is the ellipticity. 

To date several CW investigations took place and, although no signal has been detected so far, stringent upper limits on the  GW amplitude have been placed \cite{riles}.  
Each search uses a different method and is dependent on the parameter space investigated. Generally speaking the searches are divided into: \emph{targeted} or \emph{narrow-band}, when all the  source parameters (frequency, spin-down and sky position) are assumed to be accurately known, or known with a small uncertainty for the narrow-band case; \emph{directed}, which is the focus of this work, for which only the source sky position is known or barely known; and \emph{all-sky} searches where no assumptions about the source parameters are done. Latest results from O2 data are available for all-sky searches in \cite{all-skyO2}, for narrow-band searches in \cite{narrowO2} and for targeted searches in \cite{targetO2}. 

In general, in directed searches interesting sky regions or astrophysical objects are investigated, and only loose constraints on the source frequency and frequency derivatives are assumed. For this reason the parameter space covered in directed searches is wider than that of targeted and narrow-band searches, while the computational load is smaller compared to all-sky searches. The latest targets investigated in O1 directed searches include supernova remnants, globular clusters like Terzan 5 and LMXB \cite{15SNRO1,LMXBO1,O1GCTerzan5,O1EatHcasAetc}.  A previous Galactic Center CW search has been performed on two years of data from the fifth science run of LIGO \cite{GCberit}. 

In this work we consider sources potentially emitting CWs located within the inner parsecs of the Galactic Center (GC), assumed equal to the sky position of the super-massive BH Sgr A* \cite{Reid_2004}. This region could be a rich place to look for CWs, since it is likely to host several NSs, as pointed out by multiple independent lines of evidence. In a recent work  \cite{NSinGCkim} the authors report some estimates of the NS population, inferred from various observations, claiming that up to 10\% of galactic NS may occupy this central region. As already pointed out by \cite{FermiGeVBartels,FermiGeVLee,HESShooper} an existing unseen pulsar population could explain the Galactic Center $\gamma$-ray excess measured by Fermi \cite{Fermi} and by the High Energy Stereoscopic System (HESS) collaboration \cite{HESS}. Although an order of a billion of NSs is expected to exist in the Galaxy, the much smaller number of observed NSs in the galactic center region is likely related to the sensitivity limits of the surveys, as claimed by \cite{Rajwade}, due to the presence of interstellar medium along the line of sight. Note however that, although the pulsar scenario reported above is intriguing for CWs searches, the true origin of the Galactic Center $\gamma$-ray excess is still under debate \cite{Hooper2013}.
A way to overcome this limit, and to support the pulsar population hypothesis, is to look for NSs through their GW emission, since there is no interaction between the interstellar medium and GWs, and a potential CW could be detected if it is strong enough. In addition to this aspect, if we perform a CW directed search, we do not need to constrain our search to a single GW emitted frequency and we can search over a wide frequency band.

The paper is organized as follows: in Sec.\ref{sec:search} we report the search setup and the pipeline description. In Sec. \ref{section:result} we show results of the search, while upper limits are computed in \ref{sec:upper}. Sec. \ref{sec:conclusion} is left for conclusion and discussion.

\section{The search} 
\label{sec:search}
\subsection{Advanced LIGO's second observing run}
For this search we have used open data from the second observing run (O2) of the Advanced  LIGO detectors in Hanford, Washington (H) and Livingston, Louisiana (L). The run started on the 30th of November 2016 and lasted until the 25th of August 2017. The data is available at the Gravitational Wave Open Science Center webpage \cite{GWOSC,DataO1O2}.
During data taking there was a break from 2016-12-22 23:00:00 UTC to 2017-01-04 16:00 UTC, and a commissioning period for L from the 8th of May to the 26th of May, while for H it lasted from the 8th of May until the 8th of June.
Only science segments of the last version of the calibrated data \cite{calibration} have been considered; besides, poor data quality periods have been discarded from the analysis: data before the 4th of January is not considered for the L detector, while for the H detector 35 days, from mid-March to mid-April have been excluded. A third interferometer, Advanced Virgo, was running during August but, given the lower sensitivity and the significantly shorter observation time, we did not consider it in this search.

\subsection{The pipeline}
\label{subsec:pipeline}
For this work we use a new hierarchical semi-coherent directed search pipeline based on the FrequencyHough transform \cite{FrequencyHough}. We have developed this new pipeline adapting some well established concepts and procedures, such as the use of \textit{peakmaps} and Hough maps for the selection of GW candidates \cite{FFTpeakmaps,PalombaHough,FrequencyHough,FrequencyHoughmethod}, into the new Band Sampled Data (BSD) architecture, whose properties are described in \cite{BSD}. Each BSD file contains the reprocessed time strain data $h(t)$,  down-sampled to 10 Hz from the original 16 kHz strain data, under the form of a complex time series. The BSD files can be manipulated to freely choose the parameter space to investigate.

Generally speaking, the wider the parameter space the heavier the computational load is. This is the reason why hierarchical semi-coherent methods, where each chunk of data is first analyzed coherently and then incoherently combined, have been developed \cite{Brady2000}. Most often the starting point is a set of Fast-Fourier-Transform (FFT) of the calibrated data. The chunk duration, called \textit{coherence time},  is chosen short enough to keep the signal, which is subjected to Doppler and other frequency modulations, within a single frequency bin, allowing longer FFTs at lower frequencies. On the other hand, the use of longer coherence times, which increases the search sensitivity, requires higher computing power.

In order to reduce the computational load, or to use longer FFTs at fixed available computing power,  we introduce an intermediate step before the production of the peakmaps (differently to what is done in \cite{FrequencyHoughmethod}),  consisting in a partial Doppler correction.

The coherent step relies on the BSD framework and its heterodyne corrections as described in \cite{BSD}. For this purpose the Doppler demodulation described in \cite{BSD} has been modified and applied for each 1 Hz frequency band (see Appendix \ref{appendix:multidopp} for details). The incoherent step is performed using the FrequencyHough transform \cite{FrequencyHough} where the inputs, the so called \textit{peakmaps}, have been adapted to work within the BSD framework.  We remind that the FrequencyHough algorithm maps the time-frequency peaks of the peakmaps into the  frequency and spin-down (or spin-up) plane of the source.
 
\begin{figure}[h]
\centering
\includegraphics[scale=0.25,trim=0cm 3cm 0cm 0.5cm,clip]{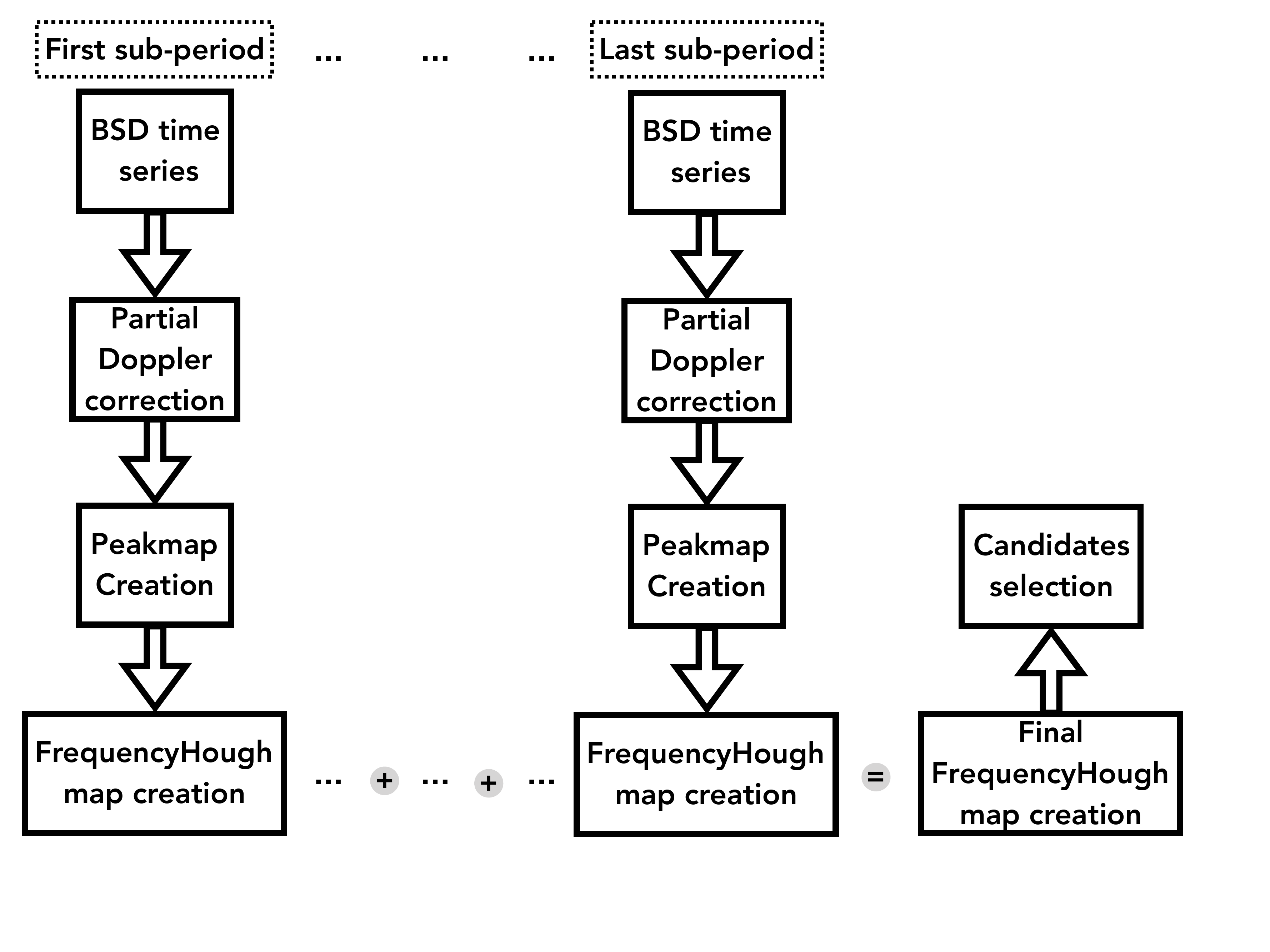}
\caption{\label{fig:scheme} Pipeline flowchart for the single detector. See text for blocks description.}
\end{figure}

In the following, we describe the steps of the pipeline and the main differences with the more general FrequencyHough method used for all-sky searches \cite{FrequencyHoughmethod}.
A scheme of the pipeline is shown in Fig. \ref{fig:scheme}.

For each BSD file covering a given 10 Hz frequency band and a run sub-period ($\sim 1$ month), the following steps are applied: 
    
1) assuming a given sky position $\vec{n}$, we partially correct the BSD complex time series using a modified version of the heterodyne used in \cite{BSD}. We repeat the correction in each 1 Hz frequency band (for details see Appendix \ref{appendix:multidopp}). Simulations show that this correction is applicable with a maximum error of 5 \% with respect to the source frequency, in a frequency band of 1 Hz.
    
2) After this partial correction, the coherence time (the FFT length) used for the peakmap can be longer, since the residual Doppler modulation will be smaller. We increase the coherence time by a factor of 4 with respect to the  FFT length computed when the signal is not corrected. 
     
3) This peakmap is the input of the FrequencyHough transform, which produces one FrequencyHough map for each BSD file. The resolution of the FrequencyHough map is given by the size of the bins of the template grid as:
    \begin{eqnarray}
\delta f_{FH} = \frac{1}{T_{coh}K_{f}}
\label{eq:deltaf}
\\
\delta \dot{f}_{FH}=\frac{1}{T_{coh}T_{obs}K_{\dot{f}}},
\label{eq:deltadf}
\end{eqnarray}
    where $T_{coh}$ is the coherence length, while $T_{obs}$ is the observational time. $K_{f}$ and $K_{\dot{f}}$ are the over-resolution factors as described in \cite{FrequencyHoughmethod}. 

4) All the produced FrequencyHough map, spanning the same frequency/spin-down bands, are summed together. We can sum up the maps since the FrequencyHough transform is a linear operation.

The final set of candidates will be selected on the total FrequencyHough map, using the same ranking procedure of \cite{FrequencyHoughmethod}. After the selection of the first level of candidates in each detector, coincidences are done between the two data-set using a coincidence distance defined as
\begin{equation}
d=\sqrt{\left(\frac{\Delta f}{\delta f_{FH}}\right)^2+\left(\frac{\Delta \dot{f}}{\delta \dot{f}_{FH}}\right)^2}
\label{Eq:distance}
\end{equation}
where $\Delta f$ and $\Delta \dot{f}$  are the differences between the parameters of the candidates of each detector. A candidate is then selected when the coincidence distance is below a given threshold distance $d_{thr}$.
Among these surviving candidates the most significant ones should be investigated in detail through a followup process (see Sec. \ref{section:result}).

\subsection{The search setup}
The total number of BSD files used for this search is 1120, spanning $N_{band} = 70$ frequency bands between 10 and 710 Hz and a spin-down range of $[-1.8 \times 10^{-9} , 3.7 \times 10^{-11}]$ Hz/s as shown in Table \ref{Tab: grid_par} where we also report the parameters that define the search grid. We remind that the frequency and spin-down bins  size, defined by Eqs. (\ref{eq:deltaf}) and (\ref{eq:deltadf}), change for each 10 Hz band. This happens because the coherence length scales with the maximum frequency of the band as  $T_{coh} \propto 1/\sqrt{f_{max}}$. 

The coherence time for the band [10 - 20] Hz is $T_{coh}=64208$ s, while it is $T_{coh}=10776$ s for the last band investigated, [700 - 710] Hz. 
The frequency resolution ranges from $1.6 \times 10^{-6}$ Hz, for the lowest frequency band, to $9.3 \times 10^{-6}$ Hz  for the band [700 - 710] Hz.  The spin-down natural resolution ranges from $3.3 \times 10^{-13}$  to $2.0 \times 10^{-12}$ Hz/s for the H detector, while for  the L detector is  $3.8 \times 10^{-13}$ Hz/s at the lowest frequency band and  $2.3 \times 10^{-12}$ Hz/s at the highest one.
For this search we have used $K_f=10$ and $K_{\dot{f}}=2$ for the frequency and spin-down bins of the FrequencyHough map.  

Concerning the sky bin, we are limiting the search to a single sky bin, hence the total number of templates will be simply the product between the number of frequency bins $N_{f}$ and the number of spin-down bins $N_{\dot{f}}$.

\begin{table}
\caption{\label{Tab: grid_par}
Frequency and spin-down ranges and main grid parameters used: $N_{f}$ is the number of frequency bins; $N_{\dot{f}}$ is the  number of spin-down bins; we used the Sgr A* as sky position for our directed search analysis. The values of $N_{f}$ and $N_{\dot{f}}$ are different for each 10 Hz band.}
\begin{ruledtabular}
\begin{tabular}{l|cdr}
Frequency  & $[10, 710]$ Hz\\
\colrule
Spin-down  & $[-1.8 \times 10^{-9} , 3.7 \times 10^{-11}]$ Hz/s\\ 
\end{tabular}
    \begin{tabular}{l|cdr}
    $N_{f}$ & $[6.42 \times 10^6, 1.08 \times 10^6]$ \\     
    \colrule
    $N_{\dot{f}}$\footnote{Since $T_{obs}$ is different in each detector, the bin size will change}  & \parbox[t]{5cm}{$[5580, 937]$for H,\\$[4860, 817]$ for L} \\ 
    \colrule
    Sky Sgr A*  &  \parbox[t]{5cm}{RA(J2000) $=17\mathrm{h}\, 45\mathrm{m}\, 40.04\mathrm{s}$,\\ Dec(J2000) $=-29^{\circ} \,00'\, 28.1"$} 
    \end{tabular}
\end{ruledtabular}
\end{table}

We perform this search pointing towards the position of Sgr A*, since we are assuming that most of the sources lie within the some parsecs from the Galactic Center. For the computation of the sky bin we use $(\lambda_{GC} , \beta_{GC} ) = (266.8517, -5.6077)^{\circ}$ in ecliptic coordinates \cite{Reid_2004}. The sky bin size not only depends on the sky position of the source, but also depends on the frequency and on the coherence time used. Indeed, the angular resolutions along the longitude and the declination will be respectively \cite{FrequencyHoughmethod}
\begin{eqnarray}
\delta \lambda = \frac{1}{N_{D}\cos \beta_{GC}}
\label{eq:deltalambda}
\\
\delta \beta=\frac{1}{N_D \sin \beta_{GC}},
\label{eq:deltadbeta}
\end{eqnarray}
where $N_D$ is the number of frequency bins affected by the Doppler effect at a given frequency which is equal to 273 for the lowest frequencies, and to 1623 for the highest frequency.
Assuming a Galactic Center distance of 8 kpc, these resolutions correspond to a sky patch centered at  $(\lambda_{GC} , \beta_{GC} )$, with a radius ranging from 150 pc (for lowest frequencies) to 25 pc (for highest ones). 

A total number of 207 jobs per detector, with a mean duration of 30 min each, run on an Intel ES-2640V4 CPU, with a total computational cost of $\sim 200$ core hours. The estimated time does not consider the BSD time production.
The total number of templates used is $2.4 \times 10^{11}$ for L and $2.7\times 10^{11}$ for H.

\section{Results}
\label{section:result}
The search produced 203961 candidates for L and 202556 for H. This number is given by the sum of all candidates selected in each of the 207 jobs per detector, where we have selected $\sim 1000$ candidates per job. Candidate selection is done through a ranking procedure on the Hough number count as in \cite{FrequencyHoughmethod}. The number of candidates chosen in each job is the result of a trade off between the need to maximize the chance of detection and the desire to followup a reasonable number of coincident candidates.
This selection is done separately for each detector. 

After the candidate selection, coincidences are done between the two datasets. 
We choose a coincidence window (see Eq. (\ref{Eq:distance})) equal to $d_{thr}=4$. This window size, supported by the analysis of data containing simulated signals, and widely discussed in \cite{FrequencyHoughmethod}, is chosen as a trade-off between the number of final candidates we are able to follow-up (which is strictly connected to the computational power available), and the need to not discard real signal candidates that can appear with slightly different parameters in the two datasets, due to noise fluctuations.

After coincidences, the surviving candidates are post-processed using first a significance threshold veto and then an additional veto consisting in the exclusion of candidates belonging to disturbed frequency regions, due to the presence of known spectral artifacts (e.g. those in \cite{covasO1O2lines}). The first selection is based on the candidate significance, given by the Critical Ratio (CR) and defined as $CR=\frac{n-Np_0}{\sqrt{Np_0(1-p_0)}}$. The CR is a measure of the statistical significance of the number count $n$ associated with the pixel of the FrequencyHough map where the candidate lies. $N$ and $p_0$ are respectively the number of FFTs and the probability of selecting a noise peak above a given peakmap threshold\footnote{The standard choice, used also in this search is $p_0=0.0755$.}. We can compute the CR threshold as in \cite{FrequencyHoughmethod}, using the false alarm probability function. In this way, the chosen CR threshold corresponds to the probability of picking an average of one false candidate over the total number of points in the parameter space. The CR threshold, which is applied separately in each detector, depends on the frequency bands and is in the range [6.00 - 6.55] for H and [5.98 - 6.53] for L.

With the choices mentioned above, we found 237 coincident candidates between the two datasets; by applying the CR threshold veto, only 9 survive. Among these 4 are due to known instrumental lines and one is produced by the presence of the hardware injection \emph{Pulsar\_10} (HI p10) \cite{Hinjections}.

The parameters of the surviving candidates are reported in Table \ref{Tab: candidates}.

\begin{table}[ht]
\caption{\label{Tab: candidates}
Candidates survived to coincidences between the two datasets. Reference time is half of the H run (13-04-17 19:42:42.95 UTC). Candidates from 5 to 9 are due to HI or known lines.}
\begin{ruledtabular}
\begin{tabular}{lccc}
cand idx & Frequency (Hz) & spin-down (Hz/s) & CR \\
\colrule 
cand 1  & 39.7583884  & -2.99 $\times 10^{-10}$ & 10.16 \\
cand 2  & 55.5978400  & -5.89 $\times 10^{-10}$ & 9.09 \\
cand 3  & 55.5982904  & -5.04 $\times 10^{-10}$ & 7.57 \\
cand 4  & 51.6780907  & -3.58 $\times 10^{-10}$ & 8.53 \\
\end{tabular}
\begin{tabular}{lcccl}
cand idx & Frequency (Hz) & spin-down (Hz/s) & CR & \\
\colrule 
cand 5  & 19.9994619 & -8.65$\times 10^{-13}$ & 13.60  & 20 Hz line \\
cand 6  & 26.3050922  &  -2.50$\times 10^{-10}$ & 7.71 & HI p10\\
cand 7 & 29.9999674 & -1.15$\times 10^{-12}$ & 15.36 & 30 Hz line\\
cand 8  & 59.9903042 &  -7.17$\times 10^{-10}$ & 8.42  & 60 Hz line\\
cand 9 & 59.9926383 & -1.09$\times 10^{-12}$ & 7.86 & 60 Hz line
\end{tabular}
\end{ruledtabular}
\end{table}
  
Interesting candidates, surviving the cleaning, overcoming the CR threshold and found in coincidence between the datasets, could be further analyzed through a followup procedure similar to the one used for surviving candidates in all-sky searches \cite{all-skyO2}. The standard idea behind a generic follow-up is to analyze the data over smaller volume, usually the same used for coincidences, using a more refined template grid and a longer coherence time after correcting the data using the frequency and the spin-down of the candidate. This stage eventually can increase the detection confidence and better estimate the candidate parameters.

\begin{figure*}[ht]
\includegraphics[scale=0.425,trim=0cm 7cm 1cm 5cm,clip]{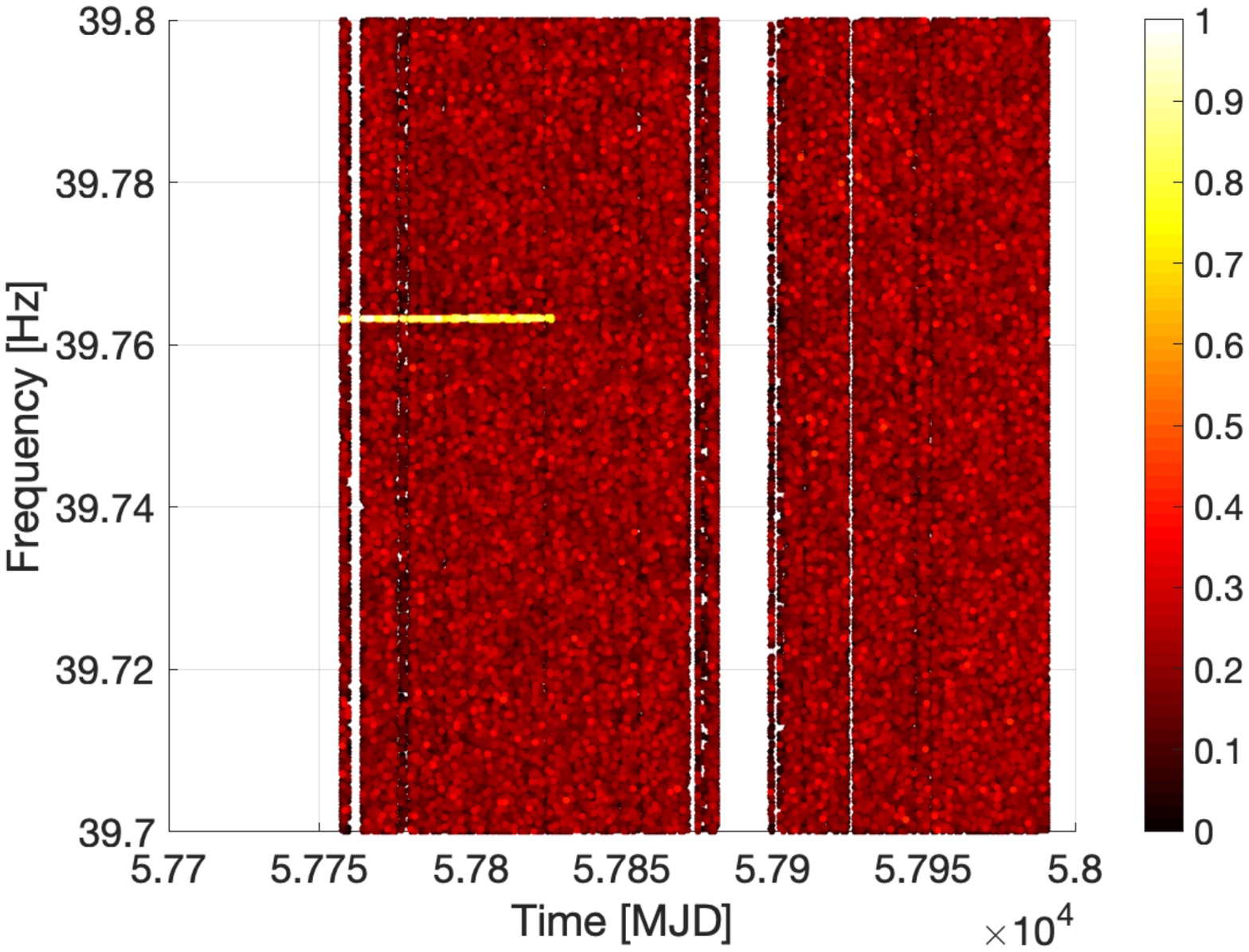}
\includegraphics[scale=0.45,trim=0cm 0cm 0cm 0cm,clip]{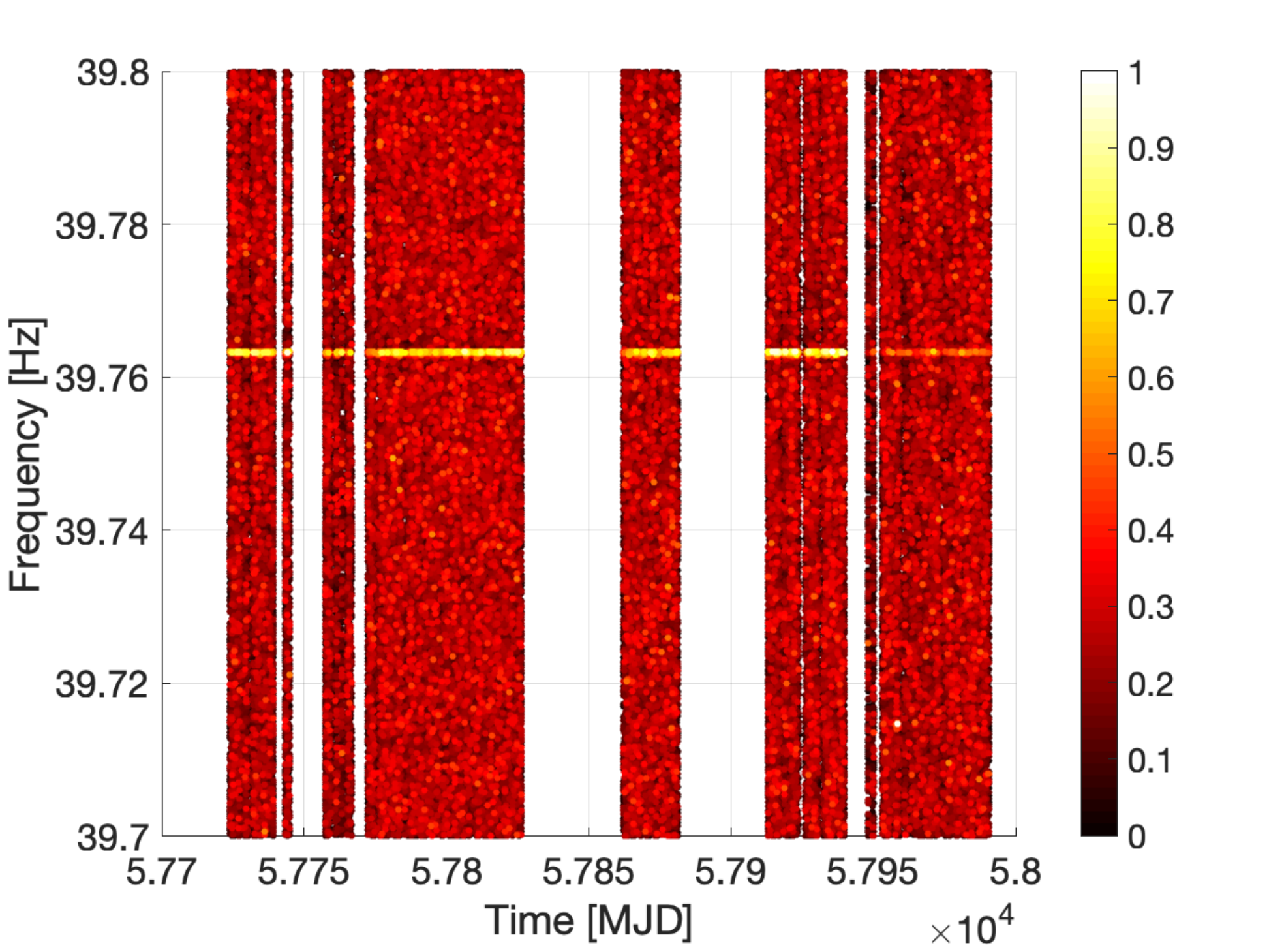}
\caption{\label{fig:example_cand} Peakmaps around the candidate at $\sim 39.76$ Hz before applying the partial-Doppler correction, for L (left) and H (right) detector data. The $x$-axis indicates time in modified Julian date (MJD), while the colorbars are the normalized values of the peaks CR.  A transient line is clearly visible in L up to $\sim 57826.5$ MJD time, while a line spanning the full run is present in H.}
\end{figure*}

Before applying the full followup procedure we can take a look to the original peakmap in a smaller frequency band around the candidate. 
As an example, for the candidate at $\sim39.76$ Hz in Fig. \ref{fig:example_cand}, we can see that there is a transient disturbance in L, lasting from the beginning of the run up to the 14th of March 2017;
while in H a line spanning the full run is visible at a frequency close to our candidate. 
In addition to visual inspection, we have found out that there was a line for L at 39.7632 Hz, coherent with auxiliary environmental monitoring channels in O1 data as reported in \cite{covasO1O2lines}. Finally, we discovered that also the rest of the candidates, show a similar transient line in L data lasting up to the 14th of March 2017. Indeed, looking at the detector logbook we have found that there was a maintenance period on that date. In particular there was a change of a power supply source which could have caused the lines to disappear \cite{log_entry}.   For this reason we strongly believe that these have been produced by non-astrophysical sources. No further followup is then needed to confirm these candidates. 

\section{upper limits}
\label{sec:upper}
Since all coincident candidates were not significant enough or they were due to spectral artifacts, we compute upper limits on the strain amplitude. 
As a first step, we compute these values on 13 trial bands of 1 Hz each, choosing those with no disturbances or hardware injected signals, we then increased the number of bands to 26. A discussion about the validity of the method, using a larger number of bands, is done in Appendix \ref{app:ulanddepth}.
To do so, in each 1 Hz band we have injected 50 signals with a given $h_0$ and computed the corresponding detection efficiency. We repeated the injections using different values of  $h_0$ in the interval [$6.6\times 10^{-27}$ , $1.3\times 10^{-24}$].  All the injections have the same sky position, equal to the Galactic Center coordinates. The frequency and spin-down are uniformly random in the bands. The polarization parameter $\eta=-\frac{\cos \iota}{1+ \cos^2 \iota}$ is generated from a uniform distribution of $\cos \iota$ between $[-1,1]$, while $\psi$ is uniformly random in the range $[-\pi/4,\pi/4]$. 

The 95\% confidence level upper limit is given by the amplitude value, $h^{95\%}_0$, such that the detection efficiency is equal to 0.95. An injection is successful when it passes the candidate selection process.
In order to get $h^{95\%}_0$, we used the following fit for the detection efficiency $D(x)$:
\begin{equation}
    \label{eqn:FH fit}
    D(x)=K(1-e^{-A_1(x-x_{min})^{A_2}})
\end{equation}
which has been used also in Eq. (5) of \cite{all-skyO1}.
The fit parameters are $A_1$ and $A_2$, while $x-x_{min}=log_{10}(\frac{h_{inj}}{h_{min}})$ where $h_{inj}$ is the injected signal strain and $h_{min}$ is the value that satisfies $D(x_{min})=0$. $K$ is a normalization factor between the maximum measured detection efficiency and the maximum of $D(x)$.

Following the approach of \cite{depth} we extend the upper limits calculation from the 26 trial bands to the full frequency band [10-710] Hz. Indeed, as discussed in \cite{depth}, the strain amplitude is proportional to $\sqrt{S_n(f)}$, which is the square root of the noise spectral density. 
For each of the 26 bands we have computed this proportionality factor, usually known as sensitivity depth, which at the end resulted almost constant over the 26 bands analyzed. 
Although we are aware that the sensitivity depth is not a constant, since there are several factors that can affect the noise level, we find out that as a first approximation, we can safely use the mean value of the computed sensitivity depths in the randomly chosen bands (see discussion in Appendix \ref{app:ulanddepth}).
For the calculation of the full $h^{95 \%}_0(f)$ upper limit curve, we have used the same noise curve $S_n(f)$ used in the FrequencyHough O2 all-sky search paper \cite{all-skyO2}. 
A detailed discussion of the validity of this procedure, compared to the usual approach used for the all-sky FrequencyHough searches, where the upper limit is computed for every 1 Hz band, is reported in \cite{Bosons2019Palomba}.
The final upper limit curve  is given in Fig. \ref{fig:ul} (see Supplemental Material). The most sensitive results are $\sim 1.4\times 10^{-25}$ for L at $161$ Hz, and $\sim 1.6\times10^{-25}$ for H at $195$ Hz with a 95\% confidence level.  The upper limits presented do not take into account the data calibration uncertainty on the amplitudes as discussed in \cite{calibration}.
\begin{figure}[h]
\centering
\includegraphics[scale=0.47,trim=0.7cm 7.5cm 1cm 8.2cm,clip]{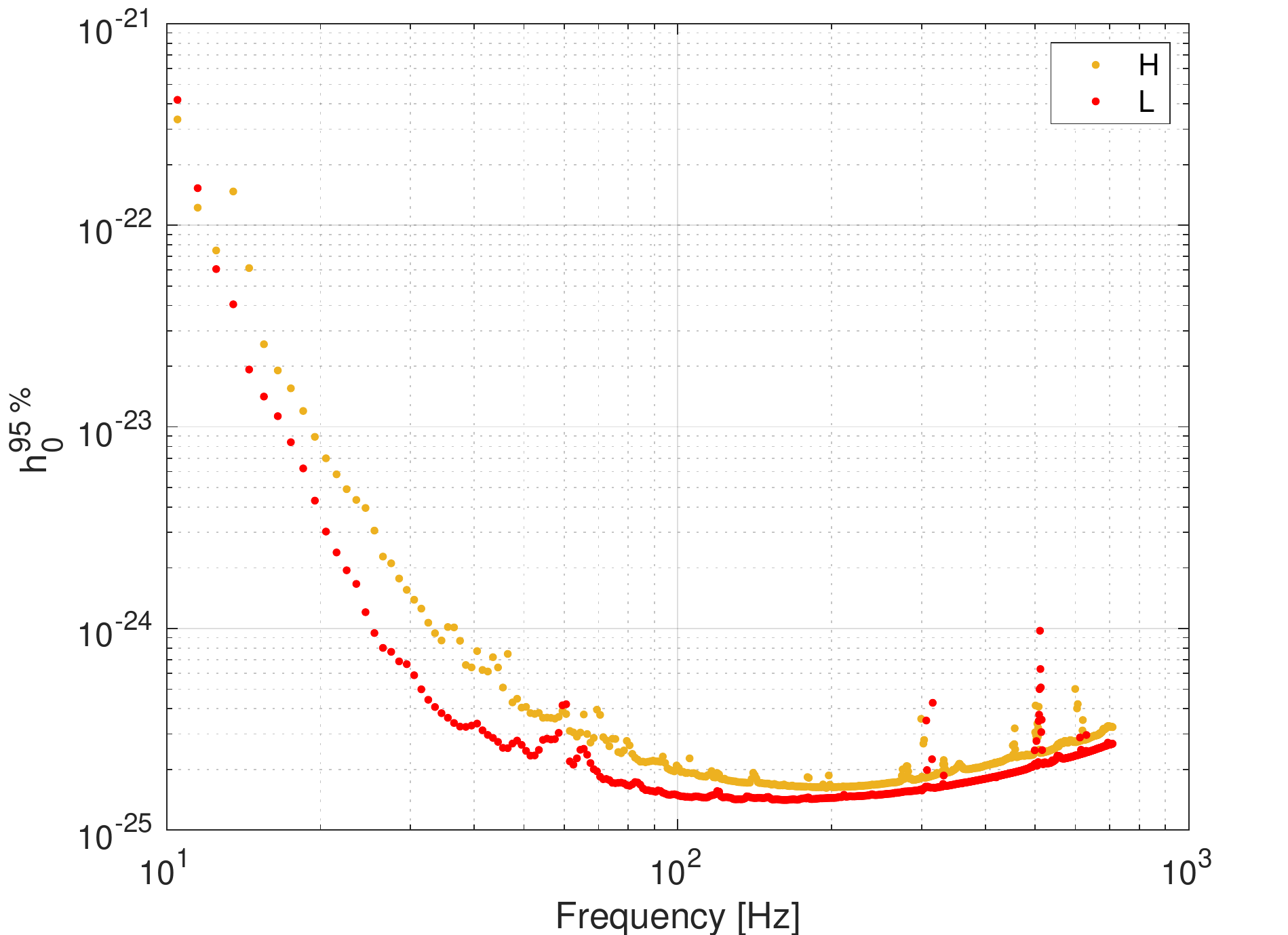}
\caption{\label{fig:ul} Upper limits of the strain amplitude at 95 \% confidence level for H and L detectors.}
\end{figure}

\begin{figure}[h]
\includegraphics[scale=0.47,trim=0.7cm 7.5cm 1cm 8.2cm,clip]{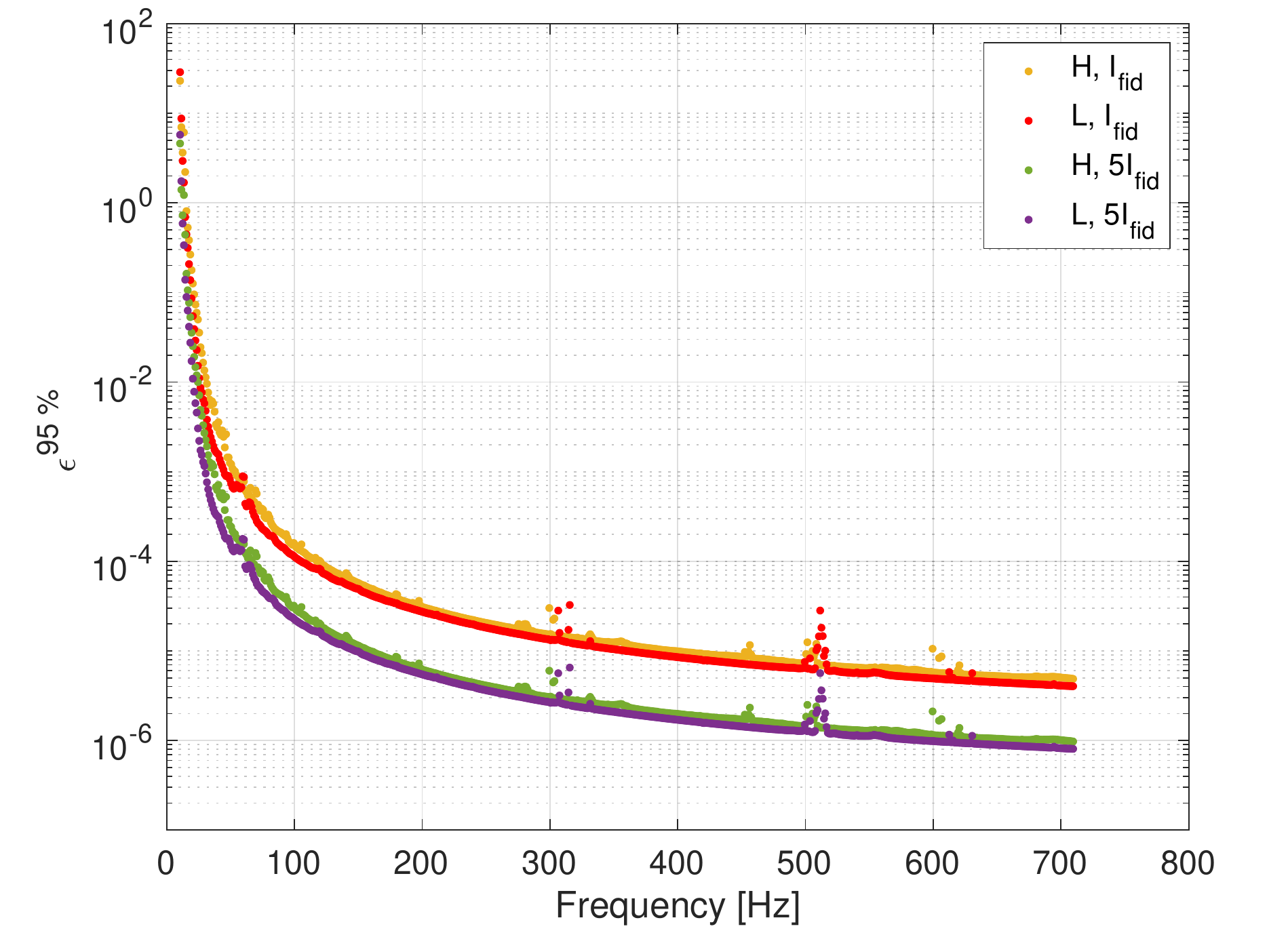}
\caption{\label{fig:ellip}Estimates of the minimum detectable ellipticity using  $I_{zz}$ equal to its fiducial value, expected for \emph{standard} NSs (yellow and red). The lower curves refer to the case of an higher moment of inertia.}
\end{figure}

Upper limits on the strain can be translated into upper limits for the ellipticity since $h_0$ and $\epsilon$ are proportional as in Eq. \ref{eqn:strain}. The results, assuming a GC distance of 8 kpc and a moment of inertia equal to the fiducial value $I_{zz}=10^{38}\,\mathrm{kg} \, \mathrm{m}^2$, are shown in Fig. \ref{fig:ellip}. We also report the ellipticity upper limits assuming a five times larger moment of inertia, which could in principle be possible for NSs with a more exotic equation of state \cite{mcdanielowen}.

The more stringent upper limit on the ellipticity is $\sim 4\times 10^{-6}$ Hz/s at the highest frequency for the L detector assuming $I_{zz}=10^{38}\,\mathrm{kg} \, \mathrm{m}^2$. This constraint is tighter if we assume higher values of the moment of inertia.

\section{conclusion}
\label{sec:conclusion}
In this work we present the first results of a directed search for CW signals from the Galactic Center in O2 data and the first results of a directed search in the band [10-500] Hz using advanced detector data.  Upper limits are comparable with O1 results in the band [500-700] Hz of \cite{O1GCTerzan5} and more stringent than those reported in the O2 all-sky search \cite{all-skyO2}. In particular \cite{O1GCTerzan5} uses longer integration times, which is a strong parameter when we want to increase the sensitivity of the search, but it also makes the search computationally heavier. On the other hand, the improvements with respect to \cite{all-skyO2} are consistent with expectation, considering the different parameter space and the longer coherence time used in this search. Indeed in \cite{O1GCTerzan5} the authors use a coherence time up to 72 hours and explore the frequencies in the range [475 - 1500] Hz and the frequency time derivatives
in the range $[-3.0; 0.1 ] \times 10^{-8}$ Hz/s; in \cite{all-skyO2} three different pipelines search for signals in the frequency band [20 - 1922] Hz and with a maximum spin-down range of $[-1; 0.2] \times 10^{-8}$ Hz/s while the pipeline that uses the maximum coherence time is the FrequencyHough, with 8192 s in the frequency band up to 128 Hz.
We have used a new directed search pipeline, developed from the Band-Sampled-Data framework. The pipeline showed an excellent computational performance in terms of computing power needed to search for a wide parameter space search. Furthermore it confirmed once again the flexibility and potentialities of the BSD framework, which can be easily adapted to many different use cases.  

From the results of this search we can exclude the presence of non-symmetric isolated spinning NS, which are emitting a CW signal bigger than our upper limits, in the Galactic Center region. In general we remind that these upper limits are valid also for particular binary systems as discussed in \cite{binariesSingh}.
These upper limits in a large frequency band [300 - 700] Hz correspond to an ellipticity smaller than $\sim 10^{-5}$, which is the maximum expected ellipticity for a normal NS \cite{mcdanielowen}.  Higher maximum ellipticities are predicted for NS with more exotic equation of state \cite{glampe}.

The LIGO and Virgo detectors have just ended the first part of the new observing run O3, started in April 2019. Both interferometers have been upgraded and the expected sensitivity is promisingly better than O2, thus increasing the detection probability. 

The pipeline described in this work could be used for the search of CW signals in O3 data, both from the Galactic Center and from other targets like supernova remnants.

\begin{acknowledgments}
This research has made use of data obtained from the Gravitational Wave Open Science Center (https://www.gw-openscience.org), a service of LIGO Laboratory, the LIGO Scientific Collaboration and the Virgo Collaboration. LIGO is funded by the U.S. National Science Foundation. Virgo is funded by the French Centre National de Recherche Scientifique (CNRS), the Italian Istituto Nazionale della Fisica Nucleare (INFN) and the Dutch Nikhef, with contributions by Polish and Hungarian institutes. 
The authors would like to acknowledge the INFN-CNAF for provision of computational resources. We would also like to thank the LIGO-Virgo CW group for useful discussions and suggestions. Ornella Juliana Piccinni would like to thank the University of Rome Sapienza for the funding provided for this work by the action ``Avvio alla Ricerca 2018". We also thank the Amaldi Research Center.
This paper carries LIGO Document Number LIGO-P1900300
\end{acknowledgments}

\appendix

\section{Doppler correction in sub-bands}
\label{appendix:multidopp}
As stated in Sec. \ref{subsec:pipeline}, the first step of the pipeline consists of applying the Doppler demodulation to the time series utilizing the heterodyne \cite{BSD}. For this search we slightly modify the implementation of the heterodyne generalizing it for the  case of Doppler correction for sources with unknown rotational parameters (in particular the emitted frequency). We call \emph{multi-Doppler} the particular partial correction algorithm described below.
When the GW emitted frequency $f_{GW}$ is known, as well as the source sky position $\vec{n}$, the phase factor which multiplies the time series is $\exp\left(i\frac{2\pi}{c}p_{\vec{n}}f_{GW}\right)$, where $p_{\vec{n}}$ is the detector position projected along the sky direction of the source. The Doppler demodulation can be implemented repeating the correction for each 1 Hz sub-band. Let us consider a single BSD file covering a 10 Hz frequency band; we extract a 1 Hz frequency sub-band in the frequency domain, getting time series of the selected sub-band. This sub-band time series is multiplied by $\exp\left(i\frac{2\pi}{c}p_{\vec{n}}f_{i}\right)$ where $f_i$ is the central frequency of the selected sub-band. We repeat the same procedure for each sub-band of 1 Hz, and the final corrected time series will be the sum of all the partially corrected sub-band time series. Simulations done with injected signals show that the correction in the sub-band is valid within a 5\% of error in the frequency (we say that the correction is valid if the signal after the correction lies in the same frequency bin where the real frequency is expected).  The residual Doppler will eventually mix with the spin-down modulation. In order to avoid losing candidates, when we do the first level selection  in the FrequencyHough map, since the error associated to the spin-down will be higher we do need to consider the over-resolution factors of $K_{\dot{f}}$ of Eq. (\ref{eq:deltadf}) when using Eq. (\ref{Eq:distance}). We have checked that if we include the over-resolution factor over the frequency it does not change the selected set of coincident candidates. 

\section{Upper limits and depths}
\label{app:ulanddepth}
The upper limits computed with the method of the sensitivity depth are only valid for bands which do not present wide disturbances. For this reason we need to state a criteria to flag a band as disturbed or undisturbed. A simple idea is to look at the power spectral distribution parameters in each band and compare these to the expected values. In particular the power spectrum noise distribution is expected to be exponential, with equal mean and standard deviation, in the case of Gaussian noise. From a practical point of view we can compute the standard deviation $\sigma$ and the mean $\mu$ of the power spectrum.  The quantity $\rho=\sigma/\mu$ is our indicator of the bad/good quality of a given band.  Given this quantity $\rho$, computed over each 1 Hz band, we flag as bad all those bands whit $\rho>\rho_{max}$, where $\rho_{max}$ is the highest value of $\rho$ among the bands chosen for the injections, which have been previously chosen as clean undisturbed bands by visual inspection. When we discard all the bands with $\rho> \rho_{max}$ we are saying that the sensitivity depth in those bands is overestimated, hence the upper limits are not valid in these bands. The list of discarded bands can be found in Tables \ref{tab:discardedH} and \ref{tab:discardedL}. With this choice we have discarded less than 10\% of the bands covered by the search, given that for H we have $\rho_{max}=1.24$ while for L $\rho_{max}=1.01$.

\begin{table}[htbp]
  \centering
  \caption{Disturbed frequency bands in H and correspondent value of $\rho$}
   \begin{ruledtabular}
    \begin{tabular}{ccc}
 idx &   Frequency (Hz) & $\rho$ \\
    \colrule
   1 & 10 - 11    & 1.67 \\
   2 & 11 - 12    & 2.61 \\
   3 & 13 - 14    & 11.82 \\
   4 & 14 - 15   & 21.76 \\
   5 & 19 - 20    & 1.39 \\
   6 & 21 - 22   & 13.84 \\
   7 & 27 - 28   & 1.25 \\
   8 & 28 - 29   & 14.83 \\
   9 & 30 - 31   & 1.30 \\
   10 & 31 - 32   & 1.28 \\
   11 & 33 - 34  & 28.66 \\
   12 & 34 - 35   & 1.36 \\
   13 & 35 - 36   & 13.60 \\
   14 & 36 - 37   & 7.24 \\
   15 & 37 - 38   & 4.20 \\
   16 & 40 - 41  & 6.62 \\
   17 & 42 - 43   & 16.94 \\
   18 & 44 - 45  & 2.10 \\
   19 & 46 - 47  & 13.59 \\
   20 & 47 - 48  & 7.48 \\
   21 & 55 - 56   & 4.31 \\
   22 & 59 - 60   & 3.47 \\
   23 & 64 - 65   & 6.25 \\
   24 & 66 - 67  & 2.06 \\
   25 & 69 - 70  & 2.04 \\
   26 & 76 - 77  & 2.42 \\
   27 & 77 - 78   & 1.81 \\
   28 & 83 - 84   & 1.50 \\
   29 & 85 - 86   & 3.01 \\
   30 & 299 - 300  & 12.00 \\
   31 & 302 - 303  & 15.13 \\
   32 & 303 - 304 & 14.97 \\
   33 & 331 - 332  & 3.34\\
   34 & 452 - 453  & 2.97\\
   35 & 486 - 487  & 1.34\\
   36 & 487 - 488  & 4.26 \\
   37 & 497 - 498  & 7.92 \\
   38 & 498 - 499  & 2.21 \\
   39 & 500 - 501  & 30.89 \\
   40 & 501 - 502   & 39.96 \\
   41 & 502 - 503  & 4.88  \\
   42 & 503 - 504  & 37.56 \\
   43 & 504 - 505  & 9.61 \\
   44 & 505 - 506  & 109.96 \\
   45 & 506 - 507  & 41.81 \\
   46 & 507 - 508  & 10.95 \\
   47 & 508 - 509  & 82.62 \\
   48 & 509 - 510  & 7.28 \\
   49 & 510 - 511  & 1.37 \\
   50 & 599 - 600  & 6.09 \\
   51 & 604 - 605  & 7.12 \\
   52 & 606 - 607  & 6.81 \\
    \end{tabular}%
    \end{ruledtabular}
  \label{tab:discardedH}%
\end{table}%

\begin{table}[htbp]
  \centering
  \caption{Disturbed frequency bands in L and correspondent value of $\rho$}
  \begin{ruledtabular}
    \begin{tabular}{ccc}
     idx &   Frequency (Hz) & $\rho$ \\
    \colrule
   1 & 10 - 11    & 1.96 \\
   2 & 11 - 12    & 1.17 \\
   3 & 12 - 13   & 1.18 \\
   4 & 13 - 14   & 18.70 \\
   5 & 14 - 15    & 8.23 \\
   6 & 15 - 16   & 2.86 \\
   7 & 16 - 17   & 2.06 \\
   8 & 17 - 18   & 38.94 \\
   9 & 18 - 19   & 1.10 \\
   10 & 19 - 20   & 10.08 \\
   11 & 20 - 21   & 1.02 \\
   12 & 21 - 22   & 1.10 \\
   13 & 22 - 23   & 1.40 \\
   14 & 23 - 24   & 3.59 \\
   15 & 24 - 25   & 1.06 \\
   16 & 25 - 26   & 1.03 \\
   17 & 26 - 27   & 1.05 \\
   18 & 27 - 28   & 1.21 \\
   19 & 28 - 29   & 1.03 \\
   20 & 30 - 31   & 1.03 \\
   21 & 31 - 32   & 1.04 \\
   22 & 33 - 34   & 1.05 \\
   23 & 35 - 36   & 2.10 \\
   24 & 40 - 41   & 1.28 \\
   25 & 42 - 43   & 1.13 \\
   26 & 59 - 60   & 1.66 \\
   27 & 60 - 61   & 1.89 \\
   28 & 119 - 120  & 2.24 \\
   29 & 199 - 200  & 1.03 \\
   30 & 306 - 307  & 15.26 \\
   31 & 307 - 308 & 13.04 \\
   32 & 314 - 315  & 7.07 \\
   33 & 315 - 316  & 10.32 \\
   34 & 331 - 332 & 2.04 \\
   35 & 495 - 496 & 1.02 \\
   36 & 499 - 500 & 52.00 \\
   37 & 500 - 501  & 1.17  \\
   38 & 503 - 504  & 11.22 \\
   39 & 507 - 508  & 1.08 \\
   40 & 508 - 509  & 26.60 \\
   41 & 509 - 510  & 62.87 \\
   42 & 510 - 511  & 48.62 \\
   43 & 511 - 512  & 77.22 \\
   44 & 512 - 513 & 9.23 \\
   45 & 513 - 514 & 87.45 \\
   46 & 514 - 515 & 7.22 \\
   47 & 515 - 516  & 32.10 \\
   48 & 516 - 517  & 9.26 \\
   49 & 517 - 518  & 3.49 \\
   50 & 518 - 519  & 3.20 \\
   51 & 519 - 520  & 10.38 \\
   52 & 527 - 528  & 1.03 \\
   53 & 528 - 529  & 1.02 \\
   54 & 612 - 613  & 4.41 \\
   55 & 615 - 616 & 2.02 \\
   56 & 629 - 630  & 1.07 \\
   57 & 630 - 631  & 2.85 \\
    \end{tabular}%
  \end{ruledtabular}
  \label{tab:discardedL}%
\end{table}%

We also want to prove that, once we determine the disturbed bands, the number of bands used to compute the depth, will only contribute to decrease the error associated to the depth, hence the final value of the depth is independent of the number of trial bands. To do so, we pick other 13 bands, computed the upper limit and the depth value in each. We get that the two mean depths, one compute from the original 13 trial bands, and the second from the second set of 13 bands is consistent and within the initial 15 \% of variance observed in the first 13 bands as shown in \ref{tab:meandepth}. 

\begin{table}[htbp]
  \centering
  \caption{Values of the depths for the bands with injections. We report the mean values computed from two different set of 13 bands.}
  \begin{ruledtabular}
    \begin{tabular}{m{0.14\textwidth}||p{0.14\textwidth} p{0.14\textwidth}}
    Frequency (Hz) & H depth ($1/\sqrt{\mathrm{Hz}}$) & L depth ($1/\sqrt{\mathrm{Hz}}$)\\
    \colrule \colrule
    32    - 33    & 49.93 & 57.26 \\
    98    - 99    & 44.33 & 52.75 \\
    100   - 101   & 48.51 & 52.71 \\
    132   - 133   & 40.10 & 53.09 \\
    182   - 183   & 44.15 & 56.13 \\
    225   - 226   & 49.28 & 57.72 \\
    326   - 327   & 47.24 & 52.47 \\
    399   - 400   & 40.96 & 51.96 \\
    425   - 426   & 42.63 & 52.90 \\
    533   - 534   & 40.30 & 47.36 \\
    627   - 628   & 44.43 & 51.02 \\
    693   - 694   & 41.74 & 50.65 \\
    694   - 695   & 42.16 & 46.78 \\
    \colrule
    Mean & 44.29 & 52.52\\
    \end{tabular}
    \begin{tabular}{m{0.14\textwidth}||p{0.14\textwidth} p{0.14\textwidth}}
    51    - 52    & 46.74 & 54.52 \\
    65    - 66    & 45.06 & 59.26 \\
    144   - 145   & 44.56 & 50.08 \\
    203   - 204   & 44.23 & 54.31 \\
    252   - 253   & 45.12 & 54.84 \\
    280   - 281   & 48.86 & 55.30 \\
    366   - 367   & 42.38 & 50.97 \\
    441   - 442   & 46.37 & 52.03 \\
    461   - 462   & 41.49 & 49.61 \\
    555   - 556   & 45.43 & 51.06 \\
    584   - 585   & 41.15 & 49.41 \\
    660   - 661   & 43.16 & 49.54 \\
    709   - 710   & 41.39 & 49.64 \\
    \colrule
    Mean & 44.30 & 52.35\\
    \end{tabular}%
    \begin{tabular}{m{0.14\textwidth}||p{0.14\textwidth} p{0.14\textwidth}}
    Total mean & 44.30 & 52.44\\
    \end{tabular}%
      
  \end{ruledtabular}
  \label{tab:meandepth}%
\end{table}%

This means that as a first approximation it is fine to consider the depth as constant even if we know that there is a small trend on the frequency, given by the different coherence time in each 10 Hz band. A way to reduce this trend (and so the variance) is to fit our depth values with a linear fit. In this way we can \textit{recalibrate} the upper limits using more accurate values for the depth in each band.


\bibliography{BSD_direc}

\begin{thebibliography}{46}%
\makeatletter
\providecommand \@ifxundefined [1]{%
 \@ifx{#1\undefined}
}%
\providecommand \@ifnum [1]{%
 \ifnum #1\expandafter \@firstoftwo
 \else \expandafter \@secondoftwo
 \fi
}%
\providecommand \@ifx [1]{%
 \ifx #1\expandafter \@firstoftwo
 \else \expandafter \@secondoftwo
 \fi
}%
\providecommand \natexlab [1]{#1}%
\providecommand \enquote  [1]{``#1''}%
\providecommand \bibnamefont  [1]{#1}%
\providecommand \bibfnamefont [1]{#1}%
\providecommand \citenamefont [1]{#1}%
\providecommand \href@noop [0]{\@secondoftwo}%
\providecommand \href [0]{\begingroup \@sanitize@url \@href}%
\providecommand \@href[1]{\@@startlink{#1}\@@href}%
\providecommand \@@href[1]{\endgroup#1\@@endlink}%
\providecommand \@sanitize@url [0]{\catcode `\\12\catcode `\$12\catcode
  `\&12\catcode `\#12\catcode `\^12\catcode `\_12\catcode `\%12\relax}%
\providecommand \@@startlink[1]{}%
\providecommand \@@endlink[0]{}%
\providecommand \url  [0]{\begingroup\@sanitize@url \@url }%
\providecommand \@url [1]{\endgroup\@href {#1}{\urlprefix }}%
\providecommand \urlprefix  [0]{URL }%
\providecommand \Eprint [0]{\href }%
\providecommand \doibase [0]{https://doi.org/}%
\providecommand \selectlanguage [0]{\@gobble}%
\providecommand \bibinfo  [0]{\@secondoftwo}%
\providecommand \bibfield  [0]{\@secondoftwo}%
\providecommand \translation [1]{[#1]}%
\providecommand \BibitemOpen [0]{}%
\providecommand \bibitemStop [0]{}%
\providecommand \bibitemNoStop [0]{.\EOS\space}%
\providecommand \EOS [0]{\spacefactor3000\relax}%
\providecommand \BibitemShut  [1]{\csname bibitem#1\endcsname}%
\let\auto@bib@innerbib\@empty
\bibitem [{\citenamefont {{J. Aasi et al. (LIGO Scientific
  Collaboration)}}(2015)}]{adligo}%
  \BibitemOpen
  \bibfield  {author} {\bibinfo {author} {\bibnamefont {{J. Aasi et al. (LIGO
  Scientific Collaboration)}}},\ }\bibfield  {title} {\bibinfo {title}
  {{Advanced LIGO}},\ }\href@noop {} {\bibfield  {journal} {\bibinfo  {journal}
  {Class. Quant. Grav.}\ }\textbf {\bibinfo {volume} {32}},\ \bibinfo {pages}
  {074001} (\bibinfo {year} {2015})}\BibitemShut {NoStop}%
\bibitem [{\citenamefont {{F. Acernese et al. (Virgo
  Collaboration)}}(2015)}]{advirgo}%
  \BibitemOpen
  \bibfield  {author} {\bibinfo {author} {\bibnamefont {{F. Acernese et al.
  (Virgo Collaboration)}}},\ }\bibfield  {title} {\bibinfo {title} {{Advanced
  Virgo: a second-generation interferometric gravitational wave detector}},\
  }\href@noop {} {\bibfield  {journal} {\bibinfo  {journal} {Class. Quant.
  Grav.}\ }\textbf {\bibinfo {volume} {32}},\ \bibinfo {pages} {024001}
  (\bibinfo {year} {2015})}\BibitemShut {NoStop}%
\bibitem [{\citenamefont {{ B. P. Abbott et al. (LIGO Scientific Collaboration
  and Virgo Collaboration}}(2019)}]{O1O2events}%
  \BibitemOpen
  \bibfield  {author} {\bibinfo {author} {\bibnamefont {{ B. P. Abbott et al.
  (LIGO Scientific Collaboration and Virgo Collaboration}}},\ }\bibfield
  {title} {\bibinfo {title} {{GWTC-1: A Gravitational-Wave Transient Catalog of
  Compact Binary Mergers Observed by LIGO and Virgo during the First and Second
  Observing Runs}},\ }\href@noop {} {\bibfield  {journal} {\bibinfo  {journal}
  {Phys. Rev. X}\ }\textbf {\bibinfo {volume} {9}},\ \bibinfo {pages} {031040}
  (\bibinfo {year} {2019})}\BibitemShut {NoStop}%
\bibitem [{\citenamefont {Hannuksela}\ \emph {et~al.}(2019)\citenamefont
  {Hannuksela}, \citenamefont {Wong}, \citenamefont {Brito}, \citenamefont
  {Berti},\ and\ \citenamefont {Li}}]{bosonsOtto}%
  \BibitemOpen
  \bibfield  {author} {\bibinfo {author} {\bibfnamefont {O.~A.}\ \bibnamefont
  {Hannuksela}}, \bibinfo {author} {\bibfnamefont {K.~W.~K.}\ \bibnamefont
  {Wong}}, \bibinfo {author} {\bibfnamefont {R.}~\bibnamefont {Brito}},
  \bibinfo {author} {\bibfnamefont {E.}~\bibnamefont {Berti}}, and\ \bibinfo
  {author} {\bibfnamefont {T.~G.~F.}\ \bibnamefont {Li}},\ }\bibfield  {title}
  {\bibinfo {title} {{Probing the existence of ultralight bosons with a single
  gravitational-wave measurement}},\ }\href
  {https://doi.org/10.1038/s41550-019-0712-4} {\bibfield  {journal} {\bibinfo
  {journal} {Nature Astronomy}\ }\textbf {\bibinfo {volume} {3}},\ \bibinfo
  {pages} {447} (\bibinfo {year} {2019})}\BibitemShut {NoStop}%
\bibitem [{\citenamefont {Arvanitaki}\ and\ \citenamefont
  {Dubovsky}(2011)}]{bosonsArvanitaki}%
  \BibitemOpen
  \bibfield  {author} {\bibinfo {author} {\bibfnamefont {A.}~\bibnamefont
  {Arvanitaki}}and\ \bibinfo {author} {\bibfnamefont {S.}~\bibnamefont
  {Dubovsky}},\ }\bibfield  {title} {\bibinfo {title} {Exploring the string
  axiverse with precision black hole physics},\ }\href
  {https://doi.org/10.1103/PhysRevD.83.044026} {\bibfield  {journal} {\bibinfo
  {journal} {Phys. Rev. D}\ }\textbf {\bibinfo {volume} {83}},\ \bibinfo
  {pages} {044026} (\bibinfo {year} {2011})}\BibitemShut {NoStop}%
\bibitem [{\citenamefont {Brito}\ \emph {et~al.}(2017)\citenamefont {Brito},
  \citenamefont {Ghosh}, \citenamefont {Barausse}, \citenamefont {Berti},
  \citenamefont {Cardoso}, \citenamefont {Dvorkin}, \citenamefont {Klein},\
  and\ \citenamefont {Pani}}]{bosonsBrito2017}%
  \BibitemOpen
  \bibfield  {author} {\bibinfo {author} {\bibfnamefont {R.}~\bibnamefont
  {Brito}}, \bibinfo {author} {\bibfnamefont {S.}~\bibnamefont {Ghosh}},
  \bibinfo {author} {\bibfnamefont {E.}~\bibnamefont {Barausse}}, \bibinfo
  {author} {\bibfnamefont {E.}~\bibnamefont {Berti}}, \bibinfo {author}
  {\bibfnamefont {V.}~\bibnamefont {Cardoso}}, \bibinfo {author} {\bibfnamefont
  {I.}~\bibnamefont {Dvorkin}}, \bibinfo {author} {\bibfnamefont
  {A.}~\bibnamefont {Klein}}, and\ \bibinfo {author} {\bibfnamefont
  {P.}~\bibnamefont {Pani}},\ }\bibfield  {title} {\bibinfo {title} {Stochastic
  and resolvable gravitational waves from ultralight bosons},\ }\href
  {https://doi.org/10.1103/PhysRevLett.119.131101} {\bibfield  {journal}
  {\bibinfo  {journal} {Phys. Rev. Lett.}\ }\textbf {\bibinfo {volume} {119}},\
  \bibinfo {pages} {131101} (\bibinfo {year} {2017})}\BibitemShut {NoStop}%
\bibitem [{\citenamefont {Riles}(2017)}]{riles}%
  \BibitemOpen
  \bibfield  {author} {\bibinfo {author} {\bibfnamefont {K.}~\bibnamefont
  {Riles}},\ }\bibfield  {title} {\bibinfo {title} {{Recent searches for
  continuous gravitational waves}},\ }\href@noop {} {\bibfield  {journal}
  {\bibinfo  {journal} {Modern Physics Letters A}\ }\textbf {\bibinfo {volume}
  {32}},\ \bibinfo {pages} {39, 1730035} (\bibinfo {year} {2017})}\BibitemShut
  {NoStop}%
\bibitem [{\citenamefont {Prix}(2009)}]{prix2009}%
  \BibitemOpen
  \bibfield  {author} {\bibinfo {author} {\bibfnamefont {R.}~\bibnamefont
  {Prix}},\ }\bibinfo {title} {Gravitational waves from spinning neutron
  stars},\ in\ \href {https://doi.org/10.1007/978-3-540-76965-1_24} {\emph
  {\bibinfo {booktitle} {Neutron Stars and Pulsars}}},\ \bibinfo {editor}
  {edited by\ \bibinfo {editor} {\bibfnamefont {W.}~\bibnamefont {Becker}}}\
  (\bibinfo  {publisher} {Springer Berlin Heidelberg},\ \bibinfo {address}
  {Berlin, Heidelberg},\ \bibinfo {year} {2009})\ pp.\ \bibinfo {pages}
  {651--685}\BibitemShut {NoStop}%
\bibitem [{\citenamefont {{P. Lasky}}(2015)}]{lasky}%
  \BibitemOpen
  \bibfield  {author} {\bibinfo {author} {\bibnamefont {{P. Lasky}}},\
  }\bibfield  {title} {\bibinfo {title} {{Gravitational Waves from Neutron
  Stars: A Review}},\ }\href@noop {} {\bibfield  {journal} {\bibinfo  {journal}
  {Pubs. Astron. Soc. Australia}\ }\textbf {\bibinfo {volume} {32}},\ \bibinfo
  {pages} {34} (\bibinfo {year} {2015})}\BibitemShut {NoStop}%
\bibitem [{\citenamefont {{K. Glampedakis and L. Gualtieri}}(2018)}]{glampe}%
  \BibitemOpen
  \bibfield  {author} {\bibinfo {author} {\bibnamefont {{K. Glampedakis and L.
  Gualtieri}}},\ }\bibfield  {title} {\bibinfo {title} {Gravitational waves
  from single neutron stars: An advanced detector era survey},\ }\href@noop {}
  {\bibfield  {journal} {\bibinfo  {journal} {The Physics and Astrophysics of
  Neutron Stars, Astrophysics and Space Science Library}\ }\textbf {\bibinfo
  {volume} {457, Springer, Cham}} (\bibinfo {year} {2018})}\BibitemShut
  {NoStop}%
\bibitem [{\citenamefont {{N. K. Johnson-McDaniel and B. J.
  Owen}}(2013)}]{mcdanielowen}%
  \BibitemOpen
  \bibfield  {author} {\bibinfo {author} {\bibnamefont {{N. K. Johnson-McDaniel
  and B. J. Owen}}},\ }\bibfield  {title} {\bibinfo {title} {{Maximum elastic
  deformations of relativistic stars}},\ }\href@noop {} {\bibfield  {journal}
  {\bibinfo  {journal} {Phys. Rev. D}\ }\textbf {\bibinfo {volume} {88}},\
  \bibinfo {pages} {044004} (\bibinfo {year} {2013})}\BibitemShut {NoStop}%
\bibitem [{\citenamefont {Owen}(2005)}]{owen2005}%
  \BibitemOpen
  \bibfield  {author} {\bibinfo {author} {\bibfnamefont {B.~J.}\ \bibnamefont
  {Owen}},\ }\bibfield  {title} {\bibinfo {title} {{Maximum Elastic
  Deformations of Compact Stars with Exotic Equations of State}},\ }\href@noop
  {} {\bibfield  {journal} {\bibinfo  {journal} {Phys. Rev. Lett.}\ }\textbf
  {\bibinfo {volume} {95}},\ \bibinfo {pages} {211101} (\bibinfo {year}
  {2005})}\BibitemShut {NoStop}%
\bibitem [{\citenamefont {{G. Woan, M. D. Pitkin, B. Haskell, D. I. Jones, and
  P. D. Lasky}}(2018)}]{Woan2019}%
  \BibitemOpen
  \bibfield  {author} {\bibinfo {author} {\bibnamefont {{G. Woan, M. D. Pitkin,
  B. Haskell, D. I. Jones, and P. D. Lasky}}},\ }\bibfield  {title} {\bibinfo
  {title} {{Evidence for a minimum ellipticity in millisecond pulsars}},\
  }\href@noop {} {\bibfield  {journal} {\bibinfo  {journal} {Astrophys. J.
  Lett.}\ }\textbf {\bibinfo {volume} {863}},\ \bibinfo {pages} {L40} (\bibinfo
  {year} {2018})}\BibitemShut {NoStop}%
\bibitem [{\citenamefont {{B. P. Abbott et al. (LIGO Scientific Collaboration
  and Virgo Collaboration)}}(2019)}]{all-skyO2}%
  \BibitemOpen
  \bibfield  {author} {\bibinfo {author} {\bibnamefont {{B. P. Abbott et al.
  (LIGO Scientific Collaboration and Virgo Collaboration)}}},\ }\bibfield
  {title} {\bibinfo {title} {{All-sky search for continuous gravitational waves
  from isolated neutron stars using Advanced LIGO O2 data}},\ }\href@noop {}
  {\bibfield  {journal} {\bibinfo  {journal} {Phys. Rev. D}\ }\textbf {\bibinfo
  {volume} {100}},\ \bibinfo {pages} {024004} (\bibinfo {year}
  {2019})}\BibitemShut {NoStop}%
\bibitem [{\citenamefont {{B. P. Abbott et al. (LIGO Scientific Collaboration
  and Virgo Collaboration}}(2019)}]{narrowO2}%
  \BibitemOpen
  \bibfield  {author} {\bibinfo {author} {\bibnamefont {{B. P. Abbott et al.
  (LIGO Scientific Collaboration and Virgo Collaboration}}},\ }\bibfield
  {title} {\bibinfo {title} {{ Narrow-band search for gravitational waves from
  known pulsars using the second LIGO observing run}},\ }\href@noop {}
  {\bibfield  {journal} {\bibinfo  {journal} {Phys. Rev. D}\ }\textbf {\bibinfo
  {volume} {99}},\ \bibinfo {pages} {122002} (\bibinfo {year}
  {2019})}\BibitemShut {NoStop}%
\bibitem [{\citenamefont {{B. P. Abbott et al. (LIGO Scientific Collaboration
  and Virgo Collaboration, radio astronomers and NICER science team
  members)}}(2019)}]{targetO2}%
  \BibitemOpen
  \bibfield  {author} {\bibinfo {author} {\bibnamefont {{B. P. Abbott et al.
  (LIGO Scientific Collaboration and Virgo Collaboration, radio astronomers and
  NICER science team members)}}},\ }\bibfield  {title} {\bibinfo {title}
  {{Searches for gravitational waves from known pulsars at two harmonics in
  2015-2017 LIGO data}},\ }\href@noop {} {\bibfield  {journal} {\bibinfo
  {journal} {Astrophys. J.}\ }\textbf {\bibinfo {volume} {879}},\ \bibinfo
  {pages} {10} (\bibinfo {year} {2019})}\BibitemShut {NoStop}%
\bibitem [{\citenamefont {{B. P. Abbott et al. (LIGO Scientific Collaboration
  and Virgo Collaboration)}}(2019)}]{15SNRO1}%
  \BibitemOpen
  \bibfield  {author} {\bibinfo {author} {\bibnamefont {{B. P. Abbott et al.
  (LIGO Scientific Collaboration and Virgo Collaboration)}}},\ }\bibfield
  {title} {\bibinfo {title} {{Searches for Continuous Gravitational Waves from
  15 Supernova Remnants and Fomalhaut b with Advanced {LIGO}}},\ }\href
  {https://doi.org/10.3847/1538-4357/ab113b} {\bibfield  {journal} {\bibinfo
  {journal} {The Astrophysical Journal}\ }\textbf {\bibinfo {volume} {875}},\
  \bibinfo {pages} {122} (\bibinfo {year} {2019})}\BibitemShut {NoStop}%
\bibitem [{\citenamefont {et~al.}(2017)}]{LMXBO1}%
  \BibitemOpen
  \bibfield  {author} {\bibinfo {author} {\bibfnamefont {B.~P.~A.}\
  \bibnamefont {et~al.}} (\bibinfo {collaboration} {LIGO Scientific
  Collaboration and Virgo Collaboration}),\ }\bibfield  {title} {\bibinfo
  {title} {{Search for gravitational waves from Scorpius X-1 in the first
  Advanced LIGO observing run with a hidden Markov model}},\ }\href
  {https://doi.org/10.1103/PhysRevD.95.122003} {\bibfield  {journal} {\bibinfo
  {journal} {Phys. Rev. D}\ }\textbf {\bibinfo {volume} {95}},\ \bibinfo
  {pages} {122003} (\bibinfo {year} {2017})}\BibitemShut {NoStop}%
\bibitem [{\citenamefont {Dergachev}\ \emph {et~al.}(2019)\citenamefont
  {Dergachev}, \citenamefont {Papa}, \citenamefont {Steltner},\ and\
  \citenamefont {Eggenstein}}]{O1GCTerzan5}%
  \BibitemOpen
  \bibfield  {author} {\bibinfo {author} {\bibfnamefont {V.}~\bibnamefont
  {Dergachev}}, \bibinfo {author} {\bibfnamefont {M.~A.}\ \bibnamefont {Papa}},
  \bibinfo {author} {\bibfnamefont {B.}~\bibnamefont {Steltner}}, and\ \bibinfo
  {author} {\bibfnamefont {H.-B.}\ \bibnamefont {Eggenstein}},\ }\bibfield
  {title} {\bibinfo {title} {{Loosely coherent search in LIGO O1 data for
  continuous gravitational waves from Terzan 5 and the Galactic Center}},\
  }\href {https://doi.org/10.1103/PhysRevD.99.084048} {\bibfield  {journal}
  {\bibinfo  {journal} {Phys. Rev. D}\ }\textbf {\bibinfo {volume} {99}},\
  \bibinfo {pages} {084048} (\bibinfo {year} {2019})}\BibitemShut {NoStop}%
\bibitem [{\citenamefont {Ming}\ \emph {et~al.}(2019)\citenamefont {Ming},
  \citenamefont {Papa}, \citenamefont {Singh}, \citenamefont {Eggenstein},
  \citenamefont {Zhu}, \citenamefont {Dergachev}, \citenamefont {Hu},
  \citenamefont {Prix}, \citenamefont {Machenschalk}, \citenamefont {Beer},
  \citenamefont {Behnke},\ and\ \citenamefont {Allen}}]{O1EatHcasAetc}%
  \BibitemOpen
  \bibfield  {author} {\bibinfo {author} {\bibfnamefont {J.}~\bibnamefont
  {Ming}}, \bibinfo {author} {\bibfnamefont {M.~A.}\ \bibnamefont {Papa}},
  \bibinfo {author} {\bibfnamefont {A.}~\bibnamefont {Singh}}, \bibinfo
  {author} {\bibfnamefont {H.-B.}\ \bibnamefont {Eggenstein}}, \bibinfo
  {author} {\bibfnamefont {S.~J.}\ \bibnamefont {Zhu}}, \bibinfo {author}
  {\bibfnamefont {V.}~\bibnamefont {Dergachev}}, \bibinfo {author}
  {\bibfnamefont {Y.}~\bibnamefont {Hu}}, \bibinfo {author} {\bibfnamefont
  {R.}~\bibnamefont {Prix}}, \bibinfo {author} {\bibfnamefont {B.}~\bibnamefont
  {Machenschalk}}, \bibinfo {author} {\bibfnamefont {C.}~\bibnamefont {Beer}},
  \bibinfo {author} {\bibfnamefont {O.}~\bibnamefont {Behnke}}, and\ \bibinfo
  {author} {\bibfnamefont {B.}~\bibnamefont {Allen}},\ }\bibfield  {title}
  {\bibinfo {title} {{Results from an Einstein@Home search for continuous
  gravitational waves from Cassiopeia A, Vela Jr., and G347.3}},\ }\href
  {https://doi.org/10.1103/PhysRevD.100.024063} {\bibfield  {journal} {\bibinfo
   {journal} {Phys. Rev. D}\ }\textbf {\bibinfo {volume} {100}},\ \bibinfo
  {pages} {024063} (\bibinfo {year} {2019})}\BibitemShut {NoStop}%
\bibitem [{\citenamefont {{Aasi, J. et al.}}(2013)}]{GCberit}%
  \BibitemOpen
  \bibfield  {author} {\bibinfo {author} {\bibnamefont {{Aasi, J. et al.}}}
  (\bibinfo {collaboration} {LIGO Scientific Collaboration and Virgo
  Collaboration}),\ }\bibfield  {title} {\bibinfo {title} {Directed search for
  continuous gravitational waves from the galactic center},\ }\href
  {https://doi.org/10.1103/PhysRevD.88.102002} {\bibfield  {journal} {\bibinfo
  {journal} {Phys. Rev. D}\ }\textbf {\bibinfo {volume} {88}},\ \bibinfo
  {pages} {102002} (\bibinfo {year} {2013})}\BibitemShut {NoStop}%
\bibitem [{\citenamefont {Reid}\ and\ \citenamefont
  {Brunthaler}(2004)}]{Reid_2004}%
  \BibitemOpen
  \bibfield  {author} {\bibinfo {author} {\bibfnamefont {M.~J.}\ \bibnamefont
  {Reid}}and\ \bibinfo {author} {\bibfnamefont {A.}~\bibnamefont
  {Brunthaler}},\ }\bibfield  {title} {\bibinfo {title} {{The Proper Motion of
  Sagittarius A{$^\ast$}. {II}. The Mass of Sagittarius A{$^\ast$}}},\ }\href
  {https://doi.org/10.1086/424960} {\bibfield  {journal} {\bibinfo  {journal}
  {The Astrophysical Journal}\ }\textbf {\bibinfo {volume} {616}},\ \bibinfo
  {pages} {872} (\bibinfo {year} {2004})}\BibitemShut {NoStop}%
\bibitem [{\citenamefont {{K. Chunglee and M. B. Davies}}(2016)}]{NSinGCkim}%
  \BibitemOpen
  \bibfield  {author} {\bibinfo {author} {\bibnamefont {{K. Chunglee and M. B.
  Davies}}},\ }\bibfield  {title} {\bibinfo {title} {{Neutron stars in the
  Galactic center }},\ }\href@noop {} {\bibfield  {journal} {\bibinfo
  {journal} {Journal of the Korean Astronomical Society}\ }\textbf {\bibinfo
  {volume} {51}},\ \bibinfo {pages} {5, 165} (\bibinfo {year}
  {2016})}\BibitemShut {NoStop}%
\bibitem [{\citenamefont {{R. Bartels, S. Krishnamurthy and C.
  Weniger}}(2016)}]{FermiGeVBartels}%
  \BibitemOpen
  \bibfield  {author} {\bibinfo {author} {\bibnamefont {{R. Bartels, S.
  Krishnamurthy and C. Weniger}}},\ }\bibfield  {title} {\bibinfo {title}
  {{Strong support for the millisecond pulsar origin of the Galactic center GeV
  excess}},\ }\href@noop {} {\bibfield  {journal} {\bibinfo  {journal} {Phys.
  Rev. Lett.}\ }\textbf {\bibinfo {volume} {116}},\ \bibinfo {pages} {051102}
  (\bibinfo {year} {2016})}\BibitemShut {NoStop}%
\bibitem [{\citenamefont {{S. K. Lee, M. Lisanti, B. R. Safdi, T. R. Slatyer
  and W. Xue}}(2016)}]{FermiGeVLee}%
  \BibitemOpen
  \bibfield  {author} {\bibinfo {author} {\bibnamefont {{S. K. Lee, M. Lisanti,
  B. R. Safdi, T. R. Slatyer and W. Xue}}},\ }\bibfield  {title} {\bibinfo
  {title} {{Evidence for Unresolved Gamma-Ray Point Sources in the Inner
  Galaxy}},\ }\href@noop {} {\bibfield  {journal} {\bibinfo  {journal} {Phys.
  Rev. Lett.}\ }\textbf {\bibinfo {volume} {116}},\ \bibinfo {pages} {051103}
  (\bibinfo {year} {2016})}\BibitemShut {NoStop}%
\bibitem [{\citenamefont {{D. Hooper, I. Cholis and T.
  Linden}}(2018)}]{HESShooper}%
  \BibitemOpen
  \bibfield  {author} {\bibinfo {author} {\bibnamefont {{D. Hooper, I. Cholis
  and T. Linden}}},\ }\bibfield  {title} {\bibinfo {title} {{TeV Gamma-Rays
  from Galactic Center Pulsars}},\ }\href@noop {} {\bibfield  {journal}
  {\bibinfo  {journal} {Physics of the Dark Universe}\ }\textbf {\bibinfo
  {volume} {21}},\ \bibinfo {pages} {40} (\bibinfo {year} {2018})}\BibitemShut
  {NoStop}%
\bibitem [{\citenamefont {{M. Ajello et al. (Fermi-LAT
  Collaboration)}}()}]{Fermi}%
  \BibitemOpen
  \bibfield  {author} {\bibinfo {author} {\bibnamefont {{M. Ajello et al.
  (Fermi-LAT Collaboration)}}},\ }\bibfield  {title} {\bibinfo {title}
  {{Characterizing the population of pulsars in the inner Galaxy with the Fermi
  Large Area Telescope}},\ }\href@noop {} {\bibfield  {journal} {\bibinfo
  {journal} {submitted to APJL}\ }\textbf {\bibinfo {volume}
  {arXiv:1705.00009}}}\BibitemShut {NoStop}%
\bibitem [{\citenamefont {{A. Abramowski et al. (H.E.S.S.
  collaboration)}}(2016)}]{HESS}%
  \BibitemOpen
  \bibfield  {author} {\bibinfo {author} {\bibnamefont {{A. Abramowski et al.
  (H.E.S.S. collaboration)}}},\ }\bibfield  {title} {\bibinfo {title}
  {{Acceleration of petaelectronvolt protons in the Galactic Centre}},\
  }\href@noop {} {\bibfield  {journal} {\bibinfo  {journal} {Nature}\ }\textbf
  {\bibinfo {volume} {531}},\ \bibinfo {pages} {476} (\bibinfo {year}
  {2016})}\BibitemShut {NoStop}%
\bibitem [{\citenamefont {{K. Rajwade, D. Lorimer and L.
  Anderson}}(2017)}]{Rajwade}%
  \BibitemOpen
  \bibfield  {author} {\bibinfo {author} {\bibnamefont {{K. Rajwade, D. Lorimer
  and L. Anderson}}},\ }\bibfield  {title} {\bibinfo {title} {{Detecting
  pulsars in the Galactic centre}},\ }\href@noop {} {\bibfield  {journal}
  {\bibinfo  {journal} {Monthly Notices of the Royal Astronomical Society}\
  }\textbf {\bibinfo {volume} {471}},\ \bibinfo {pages} {1, 730} (\bibinfo
  {year} {2017})}\BibitemShut {NoStop}%
\bibitem [{\citenamefont {{Hooper et al.}}(2013)}]{Hooper2013}%
  \BibitemOpen
  \bibfield  {author} {\bibinfo {author} {\bibnamefont {{Hooper et al.}}},\
  }\bibfield  {title} {\bibinfo {title} {{Millisecond pulsars cannot account
  for the inner Galaxy's GeV excess}},\ }\href
  {https://doi.org/10.1103/PhysRevD.88.083009} {\bibfield  {journal} {\bibinfo
  {journal} {Phys. Rev. D}\ }\textbf {\bibinfo {volume} {88}},\ \bibinfo
  {pages} {083009} (\bibinfo {year} {2013})}\BibitemShut {NoStop}%
\bibitem [{GWO()}]{GWOSC}%
  \BibitemOpen
  \href@noop {} {\bibinfo {title} {{Gravitational Wave Open Science Center}}},\
  \bibinfo {howpublished} {\url{https://doi.org/10.7935/CA75-FM95}}\BibitemShut
  {NoStop}%
\bibitem [{\citenamefont {Collaboration}\ and\ \citenamefont {the
  Virgo~Collaboration}(2019)}]{DataO1O2}%
  \BibitemOpen
  \bibfield  {author} {\bibinfo {author} {\bibfnamefont {T.~L.~S.}\
  \bibnamefont {Collaboration}}and\ \bibinfo {author} {\bibnamefont {the
  Virgo~Collaboration}},\ }\href@noop {} {\bibinfo {title} {Open data from the
  first and second observing runs of advanced ligo and advanced virgo}}
  (\bibinfo {year} {2019}),\ \Eprint {https://arxiv.org/abs/1912.11716}
  {arXiv:1912.11716 [gr-qc]} \BibitemShut {NoStop}%
\bibitem [{\citenamefont {{C. Cahillane et al.}}(2017)}]{calibration}%
  \BibitemOpen
  \bibfield  {author} {\bibinfo {author} {\bibnamefont {{C. Cahillane et
  al.}}},\ }\bibfield  {title} {\bibinfo {title} {Calibration uncertainty for
  advanced ligo’s first and second observing runs},\ }\href@noop {}
  {\bibfield  {journal} {\bibinfo  {journal} {Phys. Rev. D}\ }\textbf {\bibinfo
  {volume} {96}},\ \bibinfo {pages} {102001} (\bibinfo {year}
  {2017})}\BibitemShut {NoStop}%
\bibitem [{\citenamefont {{F. Antonucci and et al.}}(2008)}]{FrequencyHough}%
  \BibitemOpen
  \bibfield  {author} {\bibinfo {author} {\bibnamefont {{F. Antonucci and et
  al.}}},\ }\bibfield  {title} {\bibinfo {title} {{Detection of periodic
  gravitational wave sources by Hough transform in the $f$ versus $ \dot f$
  plane}},\ }\href@noop {} {\bibfield  {journal} {\bibinfo  {journal} {Class.
  Quantum Grav.}\ }\textbf {\bibinfo {volume} {25}},\ \bibinfo {pages} {184015}
  (\bibinfo {year} {2008})}\BibitemShut {NoStop}%
\bibitem [{\citenamefont {{P. Astone, S. Frasca and C.
  Palomba}}(2005)}]{FFTpeakmaps}%
  \BibitemOpen
  \bibfield  {author} {\bibinfo {author} {\bibnamefont {{P. Astone, S. Frasca
  and C. Palomba}}},\ }\bibfield  {title} {\bibinfo {title} {{The short FFT
  database and the peak map for the hierarchical search of periodic sources}},\
  }\href@noop {} {\bibfield  {journal} {\bibinfo  {journal} {Class. Quantum
  Grav}\ }\textbf {\bibinfo {volume} {22}},\ \bibinfo {pages} {1197} (\bibinfo
  {year} {2005})}\BibitemShut {NoStop}%
\bibitem [{\citenamefont {{C. Palomba et al.}}(2005)}]{PalombaHough}%
  \BibitemOpen
  \bibfield  {author} {\bibinfo {author} {\bibnamefont {{C. Palomba et al.}}},\
  }\bibfield  {title} {\bibinfo {title} {{Adaptive Hough transform for the
  search of periodic sources}},\ }\href@noop {} {\bibfield  {journal} {\bibinfo
   {journal} {Class. Quantum Grav.}\ }\textbf {\bibinfo {volume} {22}},\
  \bibinfo {pages} {S1255} (\bibinfo {year} {2005})}\BibitemShut {NoStop}%
\bibitem [{\citenamefont {{P. Astone, A. Colla, S. D’Antonio, S. Frasca, and
  C. Palomba}}(2014)}]{FrequencyHoughmethod}%
  \BibitemOpen
  \bibfield  {author} {\bibinfo {author} {\bibnamefont {{P. Astone, A. Colla,
  S. D’Antonio, S. Frasca, and C. Palomba}}},\ }\bibfield  {title} {\bibinfo
  {title} {{Method for all-sky searches of continuous gravitational wave
  signals using the frequency-Hough transform}},\ }\href@noop {} {\bibfield
  {journal} {\bibinfo  {journal} {Phys. Rev. D}\ }\textbf {\bibinfo {volume}
  {90}},\ \bibinfo {pages} {042002} (\bibinfo {year} {2014})}\BibitemShut
  {NoStop}%
\bibitem [{\citenamefont {{O. J. Piccinni et al.}}(2019)}]{BSD}%
  \BibitemOpen
  \bibfield  {author} {\bibinfo {author} {\bibnamefont {{O. J. Piccinni et
  al.}}},\ }\bibfield  {title} {\bibinfo {title} {{A new data analysis
  framework for the search of continuous gravitational wave signals}},\
  }\href@noop {} {\bibfield  {journal} {\bibinfo  {journal} {Class. Quantum
  Grav.}\ }\textbf {\bibinfo {volume} {36}},\ \bibinfo {pages} {015008}
  (\bibinfo {year} {2019})}\BibitemShut {NoStop}%
\bibitem [{\citenamefont {Brady}\ and\ \citenamefont
  {Creighton}(2000)}]{Brady2000}%
  \BibitemOpen
  \bibfield  {author} {\bibinfo {author} {\bibfnamefont {P.~R.}\ \bibnamefont
  {Brady}}and\ \bibinfo {author} {\bibfnamefont {T.}~\bibnamefont
  {Creighton}},\ }\bibfield  {title} {\bibinfo {title} {{Searching for periodic
  sources with LIGO. II. Hierarchical searches}},\ }\href
  {https://doi.org/10.1103/PhysRevD.61.082001} {\bibfield  {journal} {\bibinfo
  {journal} {Phys. Rev. D}\ }\textbf {\bibinfo {volume} {61}},\ \bibinfo
  {pages} {082001} (\bibinfo {year} {2000})}\BibitemShut {NoStop}%
\bibitem [{\citenamefont {{P. B. Covas et al.}}(2018)}]{covasO1O2lines}%
  \BibitemOpen
  \bibfield  {author} {\bibinfo {author} {\bibnamefont {{P. B. Covas et al.}}}
  (\bibinfo {collaboration} {LSC Instrument Authors}),\ }\bibfield  {title}
  {\bibinfo {title} {Identification and mitigation of narrow spectral artifacts
  that degrade searches for persistent gravitational waves in the first two
  observing runs of advanced ligo},\ }\href
  {https://doi.org/10.1103/PhysRevD.97.082002} {\bibfield  {journal} {\bibinfo
  {journal} {Phys. Rev. D}\ }\textbf {\bibinfo {volume} {97}},\ \bibinfo
  {pages} {082002} (\bibinfo {year} {2018})}\BibitemShut {NoStop}%
\bibitem [{\citenamefont {{C. Biwer et al.}}(2017)}]{Hinjections}%
  \BibitemOpen
  \bibfield  {author} {\bibinfo {author} {\bibnamefont {{C. Biwer et al.}}},\
  }\bibfield  {title} {\bibinfo {title} {{Validating gravitational-wave
  detections: The Advanced LIGO hardware injection system}},\ }\href
  {https://doi.org/10.1103/PhysRevD.95.062002} {\bibfield  {journal} {\bibinfo
  {journal} {Phys. Rev. D}\ }\textbf {\bibinfo {volume} {95}},\ \bibinfo
  {pages} {062002} (\bibinfo {year} {2017})}\BibitemShut {NoStop}%
\bibitem [{log(2017)}]{log_entry}%
  \BibitemOpen
  \href@noop {} {\bibinfo {title} {{Logbook entry n. 32262 }}},\ \bibinfo
  {howpublished}
  {\url{https://alog.ligo-la.caltech.edu/aLOG/index.php?callRep=32262}}
  (\bibinfo {year} {14 March 2017})\BibitemShut {NoStop}%
\bibitem [{\citenamefont {{B. P. Abbott et al.(LIGO Scientific Collaboration
  and Virgo Collaboration)}}(2017)}]{all-skyO1}%
  \BibitemOpen
  \bibfield  {author} {\bibinfo {author} {\bibnamefont {{B. P. Abbott et
  al.(LIGO Scientific Collaboration and Virgo Collaboration)}}},\ }\bibfield
  {title} {\bibinfo {title} {{All-sky search for periodic gravitational waves
  in the O1 LIGO data}},\ }\href@noop {} {\bibfield  {journal} {\bibinfo
  {journal} {Phys. Rev. D}\ }\textbf {\bibinfo {volume} {96}},\ \bibinfo
  {pages} {062002} (\bibinfo {year} {2017})}\BibitemShut {NoStop}%
\bibitem [{\citenamefont {{C. Dreissigacker, R. Prix, and K.
  Wette}}(2018)}]{depth}%
  \BibitemOpen
  \bibfield  {author} {\bibinfo {author} {\bibnamefont {{C. Dreissigacker, R.
  Prix, and K. Wette}}},\ }\bibfield  {title} {\bibinfo {title} {{Fast and
  accurate sensitivity estimation for continuous-gravitational-wave
  searches}},\ }\href@noop {} {\bibfield  {journal} {\bibinfo  {journal} {Phys.
  Rev. D}\ }\textbf {\bibinfo {volume} {98}},\ \bibinfo {pages} {084058}
  (\bibinfo {year} {2018})}\BibitemShut {NoStop}%
\bibitem [{\citenamefont {{C. Palomba et al.}}(2019)}]{Bosons2019Palomba}%
  \BibitemOpen
  \bibfield  {author} {\bibinfo {author} {\bibnamefont {{C. Palomba et al.}}},\
  }\bibfield  {title} {\bibinfo {title} {Direct constraints on ultra-light
  boson mass from searches for continuous gravitational waves},\ }\href
  {https://doi.org/10.1103/PhysRevLett.123.171101} {\bibfield  {journal}
  {\bibinfo  {journal} {Phys. Rev. Lett.}\ }\textbf {\bibinfo {volume} {123}},\
  \bibinfo {pages} {171101} (\bibinfo {year} {2019})}\BibitemShut {NoStop}%
\bibitem [{\citenamefont {Singh}\ \emph {et~al.}(2019)\citenamefont {Singh},
  \citenamefont {Papa},\ and\ \citenamefont {Dergachev}}]{binariesSingh}%
  \BibitemOpen
  \bibfield  {author} {\bibinfo {author} {\bibfnamefont {A.}~\bibnamefont
  {Singh}}, \bibinfo {author} {\bibfnamefont {M.~A.}\ \bibnamefont {Papa}},
  and\ \bibinfo {author} {\bibfnamefont {V.}~\bibnamefont {Dergachev}},\
  }\bibfield  {title} {\bibinfo {title} {Characterizing the sensitivity of
  isolated continuous gravitational wave searches to binary orbits},\ }\href
  {https://doi.org/10.1103/PhysRevD.100.024058} {\bibfield  {journal} {\bibinfo
   {journal} {Phys. Rev. D}\ }\textbf {\bibinfo {volume} {100}},\ \bibinfo
  {pages} {024058} (\bibinfo {year} {2019})}\BibitemShut {NoStop}%
\end{thebibliography}%
\end{document}